\documentstyle[11pt,aaspp4,psfig]{article}

\def\la{\mathrel{\hbox{\rlap{\hbox{\lower4pt\hbox{$\sim$}}}\hbox{$<$}}}}
\def\ga{\mathrel{\hbox{\rlap{\hbox{\lower4pt\hbox{$\sim$}}}\hbox{$>$}}}}

\def\etal{et\ al.}
\def\CIV{C~{\sc iv}}
\def\MgII{Mg~{\sc ii}}

\def\hMpc{$h^{-1}$~Mpc}
\def\kms{km~s$^{-1}$}

\begin{document}

\lefthead{QUASHNOCK \& STEIN}
\righthead{CLUSTERING OF QSO ABSORBERS}

\title{A New Measure of the Clustering of QSO
	Heavy--Element Absorption--Line Systems}
\author{Jean M. Quashnock}
\affil{Department of Astronomy and Astrophysics, 
University of Chicago, Chicago IL 60637}
\authoraddr{E-mail: jmq@oddjob.uchicago.edu}
\and
\author{Michael L. Stein}
\affil{Department of Statistics, 
University of Chicago, Chicago IL 60637}
\authoraddr{E-mail: stein@galton.uchicago.edu}

\received{21 August 1998}
\revised{9 November 1998}
\accepted{13 November 1998}

\slugcomment{To appear in {\em The Astrophysical Journal}}

\begin{abstract} We examine the line--of--sight clustering
of QSO heavy--element absorption--line systems,
using a new measure of clustering, called
the reduced second moment measure, $K(r)$,
that directly measures the mean over--density
of absorbers on scales $\la r$.
This measure --- while closely related to
other second--order measures such as the correlation function or the
power spectrum --- has a number of distinct statistical
properties which make possible a continuous exploration of clustering
as a function of scale.
{}From a sample of 352 \CIV\ absorbers with median
redshift $\langle z\rangle =2.2$, drawn from the spectra of 274 QSOs,
we find that the absorbers are strongly clustered on scales
from 1 to 20 \hMpc. Furthermore, there appears to be a sharp
break at 20 \hMpc, with significant
clustering on scales up to 100 \hMpc\ in excess
of that which would be expected from a smooth transition to homogeneity.
There is no evidence of clustering on scales greater than 100 \hMpc.
These results suggest that strong \CIV\ absorbers 
along a line of sight are indicators of clusters and 
possibly superclusters,
a relationship that is supported by recent observations of 
``Lyman break'' galaxies.

\end{abstract}


\section{INTRODUCTION}

In a previous series of investigations 
(\cite{VQYY96}; \cite{Quash96}; \cite{Quash98}),
the clustering properties of \CIV\ and \MgII\ absorbers have been investigated, 
using an extensive catalog of heavy--element absorption--line systems 
drawn from the literature.\footnote{Contact
D. E. Vanden Berk (danvb@astro.as.utexas.edu)
for a preliminary version of the catalog;
see York \etal\ (1991) for an earlier version.}
These authors used a line--of--sight correlation function analysis
and found evidence for strong (and evolving) 
power--law clustering on comoving
scales of 1 to 16 \hMpc\  of a form that is consistent 
with that found for galaxies and clusters at low redshift, 
and of amplitude such that absorbers are
correlated on scales of clusters of galaxies. 
Furthermore, there also appears to be
superclustering on scales of 50 to 100 \hMpc\ (Quashnock \etal\ 1996),
suggesting that these absorbers
are biased tracers of the higher--density regions of space,
and that agglomerations of strong absorbers
along a line of sight are indicators of clusters and superclusters.

This relationship is supported by recent observations of
so--called ``Lyman break'' galaxies (\cite{Steidel98})
that were found to be concentrated in coherent structures
of size $\sim 10$ \hMpc . These structures were found to contain
metal--line systems. Also, the amplitude of the correlation function 
of these Lyman break galaxies at $z=3.04$
($r_0$=$2.1\pm 0.7$ \hMpc\ [$q_0=0.5$]; \cite{Giava98})
is consistent with that found for \CIV\ absorbers 
($r_0$=$2.2$ \hMpc ; \cite{Quash98}).
While the exact relationship between high--redshift galaxies and
heavy--element absorbers is unclear,
it does appear that these systems are tracing the richer agglomerations
of the clustering network, perhaps one that is similar
to that found in detailed three--dimensional numerical investigations
of the distribution of the richest Ly$\alpha$ absorbers 
(see, e.g., \cite{Zhang98}).

Thus it is of great interest to measure and characterize the
clustering of the absorbers, over as broad a range in scale as
possible and with special attention given to the largest scales,
using the best statistical tools that are at hand.
Quashnock \etal\ (1996) were unable to relate the superclustering
found on very large ($\sim$ 100 \hMpc) scales with the power--law clustering
found later on smaller scales (\cite{Quash98}), nor locate
the approximate scale dividing these two regimes,
because they used a two--point correlation function analysis
requiring bins too large (25 \hMpc) for this purpose.

Here, we examine the line--of--sight clustering
of QSO heavy--element absorption--line systems,
using a new measure of clustering, called
the {\em reduced second moment measure}, $K(r)$, 
that directly measures the mean over--density
of absorbers on scales $\la r$. 
This measure --- while closely related to
other second--order measures such as the correlation function or the
power spectrum --- has a number of distinct statistical
properties which make possible a continuous exploration of clustering
as a function of scale.
It has been well--studied
by statisticians (\cite{Ripley88}; \cite{Badd98}) and recently
astrophysicists (\cite{Mart98}),
and several estimators have been developed for it.

The absorber catalog,
with a total of over 2200 absorbers listed over 500 QSOs,
permits exploration of clustering over a large range
in scale (from about 1 to over 100 \hMpc) and redshift 
($z$ from about 1 to 4). Ultimately, we are interested in
a three--dimensional description of the absorber distribution;
nevertheless, much of the useful information about this distribution 
lies in the one--dimensional distribution of the absorption--line systems
along the lines of sight to QSOs (see, e.g., \cite{Crotts85}).
The large number of such lines of sight makes it 
possible to make some inferences about three--dimensional clustering
from one--dimensional statistical measures.

The outline of the paper is as follows:
In \S~2 we define the reduced second moment
measure, present the estimator we have used for it,
and discuss its statistical properties. 
In \S~3 we present our results for the reduced second moment measure,
using a large sample of \CIV\ absorbers with median redshift
$\langle z\rangle =2.2$. In \S~4 we discuss the implications
of these results on our picture of absorber clustering.

\section{THE REDUCED SECOND MOMENT MEASURE}

Here we assume that the clustering
of absorbers is stationary (does not depend on time) and
homogeneous (does not depend on direction or location).
The first assumption is likely not to be strictly true,
since growth of the correlation with decreasing redshift has been detected
(\cite{Quash98}). Thus our results here are averages for our sample, 
which has a characteristic redshift given by the median
$\langle z\rangle = 2.2$.
We follow the usual convention and take the Hubble
constant, $H_0$, to be 100\,$h$\ km~s$^{-1}$~Mpc$^{-1}$
and take $q_0=0.5$ and $\Lambda=0$.

\subsection{Definitions}

Consider the process of absorber locations along some line of sight,
and let ${\cal N}$ be the mean number of absorbers per unit comoving length.
Define the {\em reduced second moment measure}, $K(r)$, as the
conditional expectation, or average ---
given that there is an absorber at $x_i$ ---
of the number of absorbers (other than the one at $x_i$ itself),
$N(x_i,r)$, that are within a comoving distance $r$ of $x_i$, 
normalized by ${\cal N}$:
\begin{equation}
K(r)=\frac{1}{\cal N} \, E\left[N(x_i,r) \mid 
{\rm absorber\; at}\; x_i\right] \;  .
\end{equation}
Because of our assumption of homogeneity,
the expected number of absorbers in equation~(1) does not depend on $x_i$.
With $q_0=0.5$ and $\Lambda=0$, the comoving distance $r$ between two
absorbers at redshifts $z_i$ and $z_j$ is 
$ r = 2c/ H_0 \times \left|{1/\sqrt{1+z_i}} - {1/\sqrt{1+z_j}}\right| $.

In terms of the two--point correlation function $\xi(r)$ 
(\cite{Peeb80}; 1993), the reduced second moment measure is given by
\begin{equation}
K(r) = 2 \int_{0}^{r} du \, \left(1+\xi(u)\right)\; .
\end{equation}
If no correlations are present, then $K(r)= 2r$.
Simply put, in this case 
the number of surrounding absorbers within 
distance $r$ of $x_i$ would not depend on the
fact that there is an absorber at $x_i$, and
would simply be equal to $2r{\cal N}$. (The factor 2 arises because
we consider distinct absorbers within a distance $r$,
or in the interval ($x_i-r$, $x_i+r$) around any given absorber.)
The quantity
$K(r)/2r \equiv 1+ \rho(r)$ is then a measure of the relative mean density
of absorbers around other absorbers, averaged over scales less than $r$.
The relative mean {\em over--density}, $\rho(r)$,
can be written in terms of the power spectrum,
$P(k)$, the Fourier transform of the correlation function $\xi(r)$,
or equivalently, in terms of the dimensionless power per logarithmic
wavenumber, $\Delta^2(k)\equiv k^3P(k)/2\pi^2$:
\begin{equation}
\rho(r) =
2 \int_{0}^{\infty}\frac{dk}{k}\, \Delta^2(k) \frac{{\rm Si}(kr)}{kr}\;  .
\end{equation}
Here ${\rm Si}(z)\equiv \int_0^z dt\, \sin(t)/t$ is the sine--integral.

Thus the reduced second moment measure, $K(r)$, is closely related to
other second--order measures such as the correlation function or the
power spectrum, and it directly measures the mean over--density
of absorbers on scales less than $r$. 
However, it has a number of distinct and desirable statistical
properties which we examine below in \S~2.3.

\subsection{Estimating $K(r)$}

Let $T_i$ be the comoving length of the $i$th line of sight, 
i.e.,  the section of the $i$th QSO spectrum which has been effectively 
searched for absorbers. 
In Figure~1, we show the cumulative distribution of 
the comoving lengths of 274 QSO lines of sight
(over an approximate redshift range $1.2<z<3.2$) 
in the Vanden Berk et al. catalog.
Almost all of the lengths are shorter than 400 \hMpc,
but the median length $\langle T\rangle =$ 350 \hMpc,
meaning that there is information on the clustering
of the absorbers on scales of 100 \hMpc\ or more.

Let $n_i$ be the number of
absorbers found in the $i$th line of sight 
at positions $x_{i1},\ldots,x_{i{n_i}}$.
If there are a total of $m$ lines of sight, 
then the total comoving length and number of absorbers are
$T=\sum_{i=1}^m T_i$ and $n=\sum_{i=1}^m n_i$, respectively.
An estimate for the mean number of absorbers per unit comoving
length is  ${\cal N}= n/T$.

{}From equation (1), a natural estimate of the reduced second
moment measure, $\hat K(r)$, is
\begin{equation}
\hat K(r) = \frac{T}{n(n-1)} \sum_{i=1}^m \sum_{j\neq j^\prime}^{n_i}
		\theta\left(r-|x_{ij}-x_{ij^\prime}|\right) \; ,
\end{equation}
where $\theta(x)$ is the Heaviside step function.
This estimate sums over pairs of absorbers that are on the same
line of sight and within distance $r$ of each other.

However, this estimator is biased low, because neighboring
absorbers that lie outside the line of sight cannot be counted.
One way to remove the bias due to edge effects is to use the 
{\em rigid motion corrected} estimator 
(\cite{Miles74}; \cite{OS81}),
which corrects for these edge effects
by weighting the summand in equation (4) 
by a factor $f(|x_{ij}-x_{ij^\prime}|)$
which depends on the separation $|x_{ij}-x_{ij^\prime}|$ relative
to the lengths $T_i$ of the lines of sight.\footnote
{Other estimators, which may have lower variance,
have been found by Stein (1993), but we
defer a discussion and treatment of these to a later work.}
This factor is the probability,
given that there is a first absorber somewhere on some line of sight,
that a second absorber of fixed separation from the first
would also be contained within the same line if sight.
We find that this probability is 
$f(|x_{ij}-x_{ij^\prime}|) = 
\sum_{l=1}^m (T_l - |x_{ij}-x_{ij^\prime}|)^+/T$
(where the $+$ superscript indicates that a summand is included
in the sum only if it is positive),
so that the edge--corrected estimator we use is 
\begin{equation}
\hat K(r) = \frac{T}{n(n-1)} \sum_{i=1}^m \sum_{j\neq j^\prime}^{n_i}
\frac{\theta\left(r-|x_{ij}-x_{ij^\prime}|\right)}{f(|x_{ij}-x_{ij^\prime}|)}
\; .
\end{equation}

\subsection{Statistical Properties}

This estimator has the following statistical property: 
$E[n(n-1)\hat K(r)/T^2]$ =
${\cal N}^2 K(r)$ {\em exactly} 
under any homogeneous and isotropic model for the absorbers.
Furthermore, while $\hat K(r)$ is not an exactly unbiased estimator
for $K(r)$, it is a consistent estimator for $K(r)$ in the
sense that $\hat K(r)$ tends to $K(r)$ in probability
as $m$ increases.

Let us contrast this estimator with the quantity $\xi_{aa}(\Delta r)$
used in Quashnock \etal\ (1996) to measure clustering.
For an interval $\Delta r=(r_1,r_2)$, $\xi_{aa}(\Delta r)$ is the
number of pairs of absorbers whose separation is in 
the interval $\Delta r$ divided
by the number of pairs that would be expected in $\Delta r$ 
if the $n$ absorbers were randomly distributed, minus~1.
This statistic has the desirable property that it tends to 0 as $m$
increases if $\xi(r)$ is identically 0 on $\Delta r$.
Furthermore, positive
values of $\xi_{aa}(\Delta r)$ indicate clustering over the range
of distances in $\Delta r$.
However, it does not provide an appropriate estimate of
$\int_{r_1}^{r_2}\xi(r)\,dr$.
In particular, $E[\xi_{aa}(\Delta r)]$ does not tend to
$\int_{r_1}^{r_2}\xi(r)\,dr$ as $m$ increases.
For example, if all lines of sight were of equal length $T_1$,
it is possible to show that as $m\to\infty$,
\begin{equation}
\xi_{aa}(\Delta r)\to {\int_{r_1}^{r_2}(T_1-u)\xi(u)\,du\over
                       T_1(r_2-r_1)-{1\over 2}(r_2^2-r_1^2)}
\quad\hbox{in probability.}
\end{equation}
The fact that this limit generally depends on $T_1$ is undesirable
for purposes of obtaining a quantitative assessment of the clustering
of absorbers.
When $\xi$ is nearly constant on $\Delta r$,
the limit in equation~(6) is approximately 
$\xi\big((r_1+r_2)/2\big)$, as one would hope.
Unfortunately, the moderate size of this data set requires the use
of rather wide bins, and Quashnock \etal\ (1996)
use values of $r_2-r_1$ of 25 \hMpc\ and greater.
Using the relationship between $\xi$ and $K$ in equation~(2), we can
easily obtain a consistent estimator of $\int_{r_1}^{r_2}\xi(r)\,dr$
as $m$ increases.
Specifically,
$[\hat K(r_2)-\hat K(r_1)]/2$ converges in probability to
$\int_{r_1}^{r_2}\xi(r)\,dr$ as $m$ increases.

By examining $\hat K(r)$ as a function of $r$, we can make 
a {\em continuous} exploration of clustering as a function of scale,
without the binning required when using a correlation function analysis. 
In particular, this permits a more detailed examination
of the relationship between small--scale clustering (\cite{Quash98})
and the superclustering found by Quashnock \etal\ (1996).
The reduced second moment measure estimator (eq.~[5])
is easy to compute and has well--understood statistical properties.

\section{RESULTS}

We have used equation (5) to estimate 
the reduced second moment measure, $\hat K(r)$,
for 274 QSO lines of sight,
obtained from the Vanden Berk et al. catalog.
A total of 352 \CIV\ absorbers have been selected from this
heterogeneous catalog, using selection criteria 
(Quashnock \etal\ 1996; \cite{Quash98}) designed to obtain
as homogeneous a data set as possible.
We refer the reader to these papers for a detailed description
of the selection criteria.

In Figure~2, we show  our results for the quantity
$\hat K(r)/2r \equiv 1 + \hat\rho(r)$ {\em (solid line)} for this sample. 
This quantity has expectation value unity,
if there is no clustering of absorbers along lines of sight (see eq. [2]). 
We have constructed 1000 data sets of 352 absorbers uniformly distributed 
along the 274 QSO lines of sight, these lines having the same distribution
of comoving lengths as in our actual data sample (see Fig.~1).
The 95\% region of variation of $\hat K(r)/2r$  
for these 1000 simulated data sets,
about the expectation value of unity,
is also shown in Figure~2 {\em (dashed lines)}.
Our estimated value of $\hat K(r)/2r$ for the \CIV\ absorber data set
is much greater than the upper limit of this band,
for values of $r$ between 1 and 20 \hMpc .
For example, for $r=10$ \hMpc\ our estimate is more than 12$\sigma$
above unity,
meaning that a simulated
data set with uniformly distributed absorbers would essentially
never have a $\hat\rho$ as large as is measured.
Thus \CIV\ absorbers cluster significantly on these scales.

We compare these results for  $\hat K(r)/2r$ with those
of Quashnock \& Vanden Berk (1998) --- who found that
the correlation function of \CIV\ absorbers on scales of
1 to 16 \hMpc\ is consistent with a power law of the
form $\xi(r)=(r_0/r)^\gamma$, with $r_0=3.4$ \hMpc\
and $\gamma=1.75$ --- by substituting this form of the
correlation function into equation (2).
In Figure~2 {\em (light line)}, we show the value of $K(r)/2r$
if absorbers have this power--law correlation function.\footnote
{Since we have no information about $\xi(r)$ on scales smaller than 1 \hMpc\ 
(\cite{Quash98}), eq.~[2] implies that $K(r)$ in this 
power--law case is determined only to within an additive constant.
Here, we have fixed $K(r)$ to its measured value at $r$ = 5 \hMpc.}
This form of clustering appears to describe the estimated reduced
second moment measure $\hat K(r)$ reasonably well, 
out to scales $r\sim $ 20 \hMpc;
afterwards, there appears to be a break in the form of $\hat K(r)/2r$.

We have investigated the significance of this excess by
examining the quantity $[\hat K(r)-\hat K(20)]/[2(r-20)] =
[\int_{20}^r \xi(u)\, du]/(r-20) \equiv \Delta_{20}(r)$,
shown in panel~{\em a} of Figure~3 {\em (solid line)} 
for the same sample of \CIV\ absorbers as in Figure~2.
{}From equation (2), $\Delta_{20}(r)$ also has expectation value of unity,
if the correlation function is zero on scales greater than 20 \hMpc .
{}From Figure~3, it appears that $\Delta_{20}(r)$ is greater than unity
on scales $r\ga $ 30 \hMpc .
We have estimated the error in the estimate of $\Delta_{20}(r)$  
by a bootstrap resampling
method in which we randomly pick 274 QSO lines of sight from the
actual data sample, with replacement, i.e., allowing for the same
line of sight to be picked multiple times
(see \cite{ET93}, or \cite{DH97}, for a review of
bootstrap methods for estimating errors). This method ensures
that the distribution of lengths of the resampled data sets
is the same as that of the actual sample (Fig.~1).
In panel~{\em a} of Figure~3 {\em (dashed lines)},
we also show the bootstrap--estimated 95\%
pointwise (i.e., for each value of $r$)
confidence region for $\Delta_{20}(r)$. While there is some uncertainty 
in the estimate of this quantity, 
it does appear that there is significant excess
clustering on scales $r \ga$ 30 \hMpc .
For example, when resampling data sets by the bootstrap method,
$\Delta_{20}$(50~\hMpc) is greater than unity 99.998\% of the time.

The bootstrap procedure for obtaining confidence intervals
we have employed here has the desirable property that its
validity does not require any special assumptions about the
nature of the absorber distribution along a line of sight.
More specifically, because it uses lines of sight as the sampling
unit in the resampling scheme, it only requires that the location
of absorbers on {\em different} lines of sight are independent.
Since the majority of  lines of sight 
are not within 100 \hMpc\ of any other line of sight,
this independence assumption is reasonable.
By using a resampling in which groups of lines of sight are resampled
rather than individual lines, we believe it should be possible
to detect if this independence assumption is appropriate.
We plan to investigate this possibility and other refinements
of the bootstrapping procedure in future work.

The procedures used by Quashnock \& Vanden Berk (1998)
and Quashnock \etal\ (1996) also assume that absorber
locations on different lines of sight are independent.
In addition, they both make use of further approximations about
the absorber location process within a line of sight.
Quashnock \& Vanden Berk (1998)
obtain approximate confidence intervals
for the line--of--sight correlation
function up to distances of 16 \hMpc\ by assuming
that every pair of points whose separation is in the interval
$\Delta r$ is independent of every other such pair.
This assumption may be a good approximation when $r_2-r_1$ is
small, although simulations in Stoyan, Bertram, \& Wendrock (1993)
suggest that such an assumption may often lead to overoptimistic
confidence intervals.
On larger scales, for which Quashnock \etal\ (1996)
have used fairly wide bins (greater than 25 \hMpc\ wide),
assuming independence between pairs with distance in $\Delta r$
may be problematic.
There, the confidence intervals are based on assuming that one can
ignore correlations beyond second--order in absorber locations
along a line of sight.
While we have no evidence that such an assumption is wrong,
the bootstrapping procedure we employ is valid whether or not this
assumption is reasonable.

We have also searched for clustering on scales greater than 50, 100,
and 150 \hMpc, by examining the quantities $\Delta_{50}(r)$, 
$\Delta_{100}(r)$, and $\Delta_{150}(r)$, shown {\em (solid lines)} 
in panels {\em b}, {\em c}, and {\em d}, respectively, of Figure~3.
Again we also show the bootstrap--estimated 95\%
confidence region for each quantity {\em (dashed lines)}.
We find that $\Delta_{50}(r)$ is significantly greater than unity, 
for scales $r > $ 50 \hMpc, meaning that
there is significant clustering on those scales.
Namely, when resampling data sets by the bootstrap method,
$\Delta_{50}$(100~\hMpc) is greater than unity 99.91\% of the time.
However, $\Delta_{150}(r)$ is statistically consistent with unity
for all $r>$ 150 \hMpc, and $\Delta_{100}(r)$ is consistent
with unity everywhere except (marginally) for $r\sim$ 200 \hMpc.
This supports the conclusion of Quashnock et al. (1996) that at present
there is no significant evidence for clustering of absorbers on scales
greater than 100 \hMpc.

\section{DISCUSSION}

We have demonstrated that the line--of--sight clustering
of QSO heavy--element absorption--line systems can be examined
using a new measure of clustering, called
the reduced second moment measure, $K(r)$, 
that directly measures the mean over--density
of absorbers on scales $\la r$. By estimating $K(r)$,
we find that the absorbers are strongly clustered on scales
from 1 to 20 \hMpc, in a manner that is consistent
with a power--law correlation function of the form 
found by Quashnock \& Vanden Berk (1998).
The form and amplitude of this clustering strongly suggests that
the absorbers are tracing the large--scale structure
seen in the distribution of galaxies and clusters.

However, because we have only examined the clustering of absorbers
in one dimension, along the line of sight, there remains
the possibility that some or all of the excess clumping is due
to velocity effects, i.e., groups of component absorbers spread out
in redshift due to velocity dispersion.
(Note that at redshift $\langle z\rangle = 2.2$, 1 \hMpc\ corresponds
to velocity differences $\Delta v$ = 180 \kms\ in the rest frame:
The flattening in $1 + \hat\rho(r)$ seen near 1 \hMpc\ in Figure~2
may be due to velocity dispersion, as well as limited spectral resolution.)
This has been argued by
Crotts, Burles, \& Tytler (1997), who explore the spatial clustering
of \CIV\ systems along adjacent lines of sight, and claim that it is
significantly weaker than clustering along a line of sight.
Quashnock \& Vanden Berk (1998) have shown that,
whether due to peculiar motions inside clusters, or to actual  spatial
clustering on megaparsec scales, that the scale, the form, and the amplitude
of the clustering are all indicative of an association of strong absorbers
with clusters. Such an association is also supported by observations
of  ``Lyman break'' galaxies (see \S\ 1).

In Figure~2, there is a sharp
break in the form of $\hat K(r)$ at 20 \hMpc .
It thus appears that (for $q_0=0.5$) this is the scale
marking the boundary of power--law clustering on smaller scales.
Using the reduced second moment measure has permitted an
approximate determination of this break.
From Figure~3 (panels {\em a}--{\em d}) there is evidence
for clustering on scales of up to 100 \hMpc\ ---
but not on larger scales --- in excess
of that which would be expected from a smooth transition to homogeneity.

One possible interpretation of this excess is that it is due
to superclustering on scales of 50 to 100 \hMpc\ 
(\cite{HHW89}; \cite{Dinshaw96}; \cite{Williger96}; 
Quashnock \etal\ 1996, and references therein),
much like what is seen locally in the distribution of galaxies:
If true, this means that these absorbers
are biased tracers of the higher--density regions of space,
and that agglomerations of strong absorbers
along a line of sight are indicators of clusters and superclusters.

However, Richards \etal\ (1999) have recently claimed that there may
be evidence in the data catalogs that the number of \CIV\ absorbers
along the line of sight depends on the intrinsic properties of the QSO.
These authors argue that there may be a significant
contamination of true intervening systems along the line of sight
by absorbers that are actually associated with the QSO,
and that such a contamination may extend to relative velocities
as great as 75000~\kms\ from the QSO.
In this work, we have adopted the standard
cutoff and excluded absorbers that are closer than 5000 \kms\
to the QSO (\cite{Foltz88};
this corresponds to comoving distances of about 30 \hMpc\ in this work).

It is possible that such a contamination is present in the
large--scale excess $\hat\rho(r)$ in Figure~2.
A more detailed analysis of this possible effect
will require an indicator 
capable of distinguishing, at least statistically, associated
absorption--line systems from true intervening ones.
While the exact nature of this large--scale excess is still uncertain,
its existence on scales of 20 \hMpc\ has been unambiguously
revealed by the present analysis.

\acknowledgments
This work was supported in part 
by NASA grant NAG~5-4406 and NSF grant DMS~97-09696 (J.~M.~Q.),
and by NSF grant DMS~95-04470 (M.~L.~S.).

\clearpage

\begin{figure*}[t]
\hbox{\hskip 0.0truein
\centerline{\psfig{file=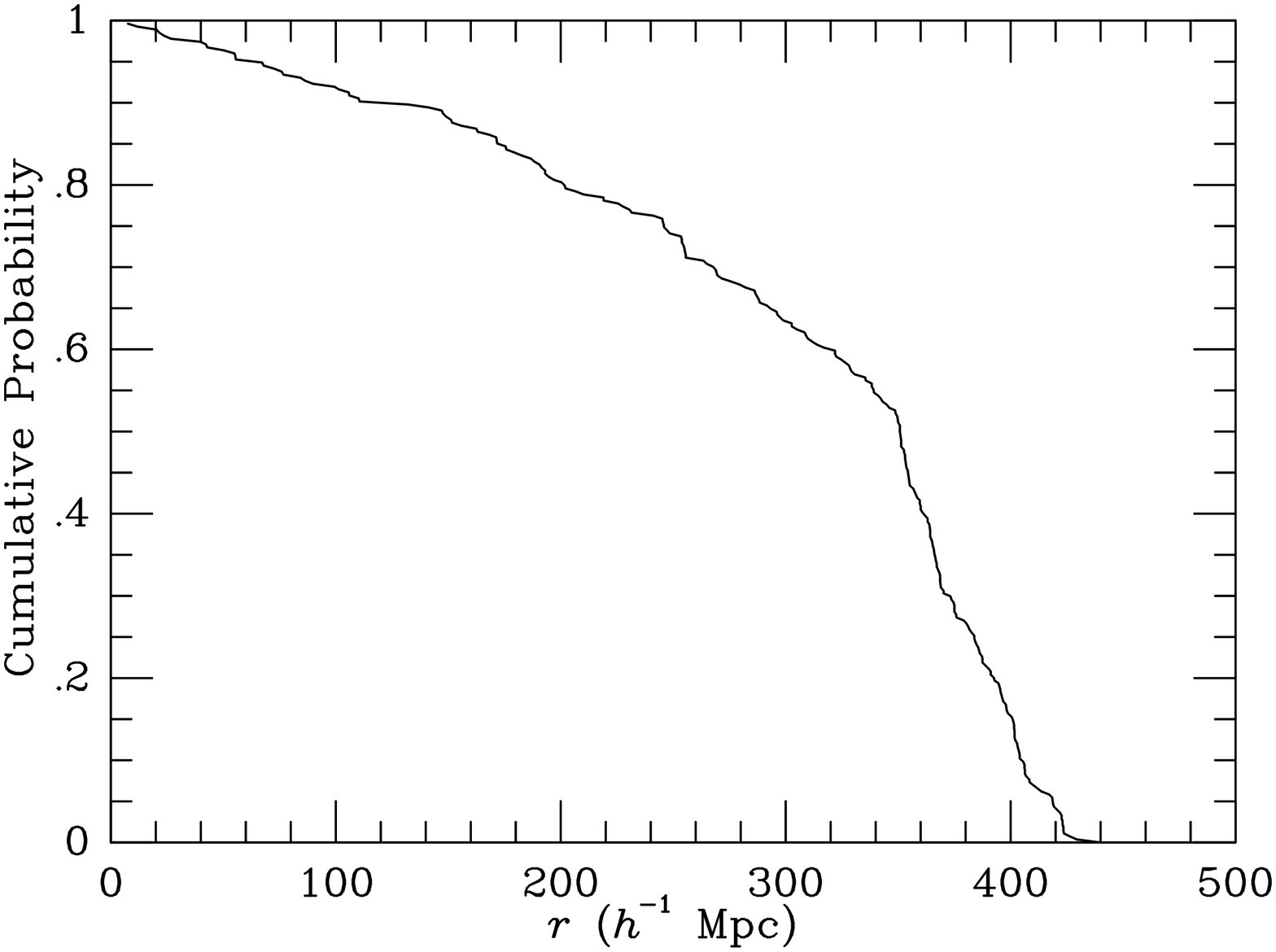,width=6.0truein,angle=-90}\hfil}}
\caption[fig1]{Cumulative distribution of 
the comoving lengths of the 274 QSO lines of sight,
containing a total of 352 \CIV\ absorbers
obtained from the Vanden Berk et al. catalog.
The median length $\langle T\rangle =$ 350 \hMpc\ ($q_0=0.5$).}
\end{figure*}

\clearpage 

\begin{figure*}[t]
\hbox{\hskip 0.0truein
\centerline{\psfig{file=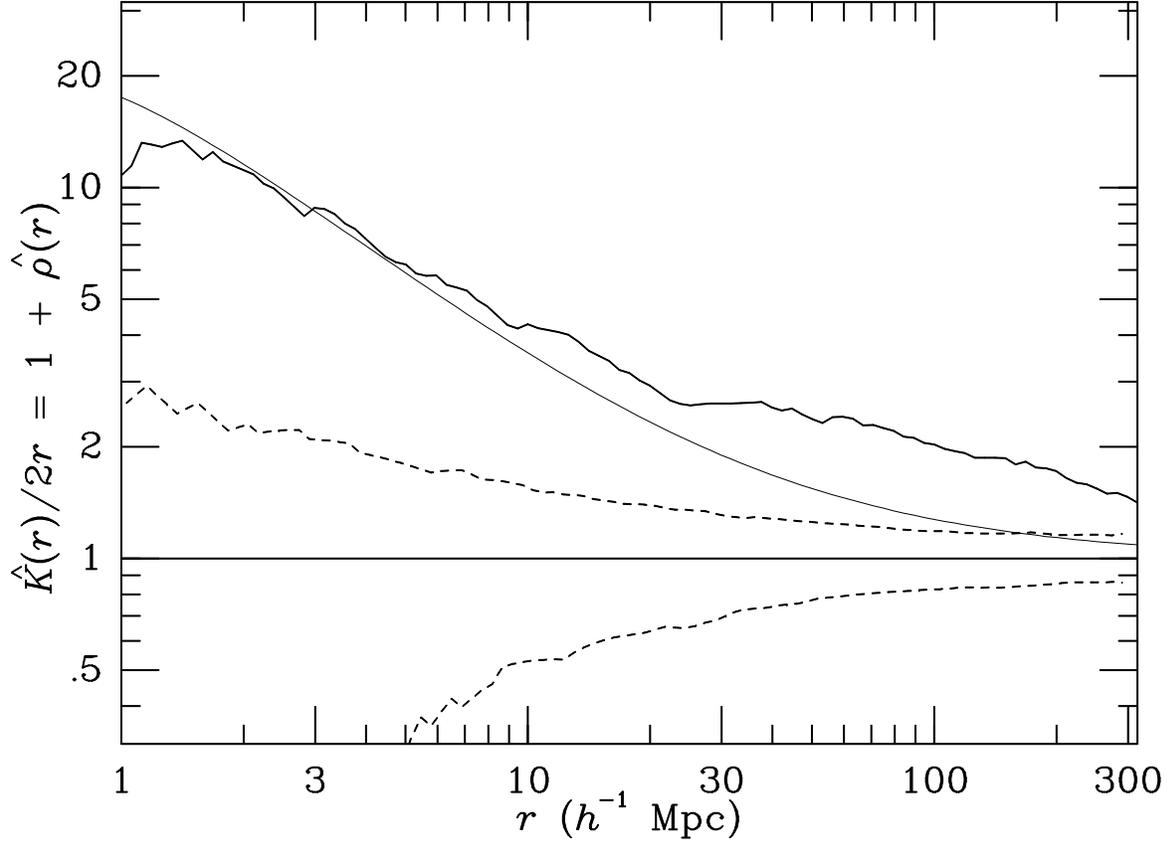,width=6.0truein,angle=-90}\hfil}}
\caption[fig2]{Estimate of the reduced second moment measure, 
$\hat K(r)$, divided by $2r$ {\em (solid line)},
for the 274 QSO lines of sight,
containing a total of 352 \CIV\ absorbers
obtained from the Vanden Berk et al. catalog.
For comparison, we show $K(r)/2r$ for absorbers that have a power--law
correlation function, $\xi(r)=(r_0/r)^\gamma$, with $r_0=3.4$ \hMpc\
and $\gamma=1.75$ ({\em light line}; see text).
Also shown is the 95\% region of variation 
of $\hat K(r)/2r$ for 1000 simulated data sets of unclustered absorbers, 
about the expectation value of unity {\em (dashed lines)}.}
\end{figure*}

\clearpage

\begin{figure*}[t]
\hbox{\hskip 0.0truein
\centerline{\psfig{file=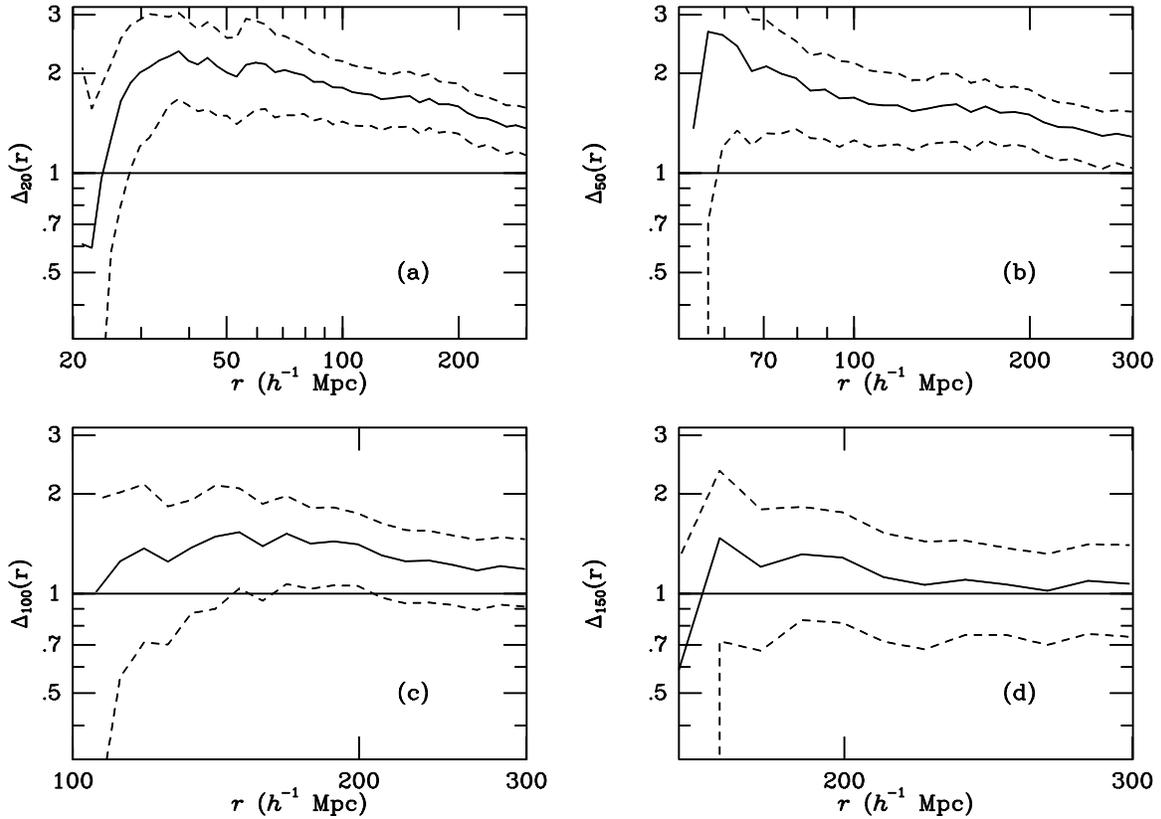,width=6.0truein,angle=-90}\hfil}}
\caption[fig3]{Clustering measure {\em (solid lines)} on scales greater than: 
(a) 20 \hMpc, $\Delta_{20}(r)$;
(b) 50 \hMpc, $\Delta_{50}(r)$;
(c) 100 \hMpc, $\Delta_{100}(r)$;
(d) 150 \hMpc, $\Delta_{150}(r)$;
for the same sample of \CIV\ absorbers as in Fig.~2.
Also shown are the bootstrap--estimated 95\%
pointwise confidence regions for each quantity ({\em dashed lines}; see text).}
\end{figure*}

\end{document}